\documentclass[11pt,hidelinks]{article}

\usepackage{amssymb, amsmath, amsthm, amsfonts, enumitem, latexsym, tikz, color, hyperref, soul}
\usepackage[margin=1in,paperwidth=8.5in,paperheight=11in]{geometry} 
\usepackage[nameinlink]{cleveref}

\usetikzlibrary{fit} 
\tikzstyle{pt}=[circle, fill=black, inner sep=2pt]

\newtheorem{theorem}{Theorem}[section]
\newtheorem*{definition*}{Definition}
\newtheorem{corollary}[theorem]{Corollary}
\newtheorem{proposition}[theorem]{Proposition}
\newtheorem{lemma}[theorem]{Lemma}
\theoremstyle{remark} \newtheorem*{remark*}{Remark}

\newcommand{\eps}{\varepsilon}

\DeclareMathOperator{\poly}{poly}
\DeclareMathOperator{\quasipoly}{quasipoly}
\DeclareMathOperator{\N}{{N}}  
\DeclareMathOperator{\mcm}{MCM}

\date{}
\begin{document}
\title{Online matching games in bipartite expanders: Hall-type results and applications to non-blocking connectors}

\author{Bruno Bauwens\thanks{National Research University Higher School of Economics, Faculty of Computer Science, Moscow, Russia;
This work/article is an output of a research project (HSE-BR-2025-024) implemented as part of the Basic Research Program at HSE University. 
}
\and
Marius Zimand\thanks{Department of Computer and Information Sciences, Towson University, Baltimore, MD.}}

\maketitle

\begin{abstract}
 The classical Hall Theorem shows the equivalence between \emph{offline} matching in bipartite graphs and a certain expansion property. We show that lossless expansion implies efficient \emph{online} matching. Online matching is defined by the following game: an opponent switches {\em on} and {\em off} nodes on the left side and, at any moment, at most $K$ nodes may be on. 
  Each time a node is turned on, it must be irrevocably matched with one of its neighbors.  
  A bipartite graph has $e$-expansion up to $K$ if every set $S$ of at most $K$ left nodes has at least $e\#S$ neighbors. 
  If all left nodes have degree $D$ and $e$ is close to $D$, then the graph is a lossless expander. 
   We show that lossless expanders allow for a polynomial-time strategy in the above game, and, furthermore, a slight modification of such a graph allows a strategy running in time $O(D \log N)$, where $N$ is the number of left nodes. 
 
  From this, improved non-blocking connectors are obtained. These are graphs with a set of input and a set of output nodes. In online fashion, requests of the form (input node, output node) are given, and each time, a path connecting the two nodes must be constructed that is vertex disjoint from paths constructed for previous requests. We obtain constant depth $N$-connectors of optimal size with efficient path-finding algorithms. For non-explicit connectors, the run time is poly$(\log N)$ and the size is optimal within $\poly(\log N)$ factor. For explicit connectors, both the runtime and optimality factor are $\exp(O(\log^2 \log N))$. Previous path-finding algorithms for such connectors were double-exponentially slower. 

 In the companion papers~\cite{companion-DS} and~\cite{companion-construction}, we consider some variants of the above online matching game and present several applications to data structures and minimum descriptions.
\end{abstract}

\section{Introduction}\label{s:intro}

A bipartite graph has {\em offline matching} up to~$K$ elements if every set of $K$ left nodes can be covered by $K$ pairwise disjoint edges.
A graph has {\em $e$-expansion} up to~$K$ if every subset $S$ with at most $K$ left nodes has at least $e \cdot \# S$ right neighbors.
Clearly, a graph with offline matching up to~$K$ has 1-expansion up to~$K$. 
Hall's theorem states that the two assertions are actually equivalent. Moreover, it is known how to construct the matchings efficiently (for example, by maximum matching algorithms from~\cite{hop-kar:j:matchbipartite} or~\cite{chen-et-al:t:maxflowlintime}).

 Is there anything similar for \emph{online matching}? In \cite[proposition 1]{fel-fri-pip:j:networks}, it is shown that $4$-expansion implies online matching, but the matching algorithm requires time exponential in the number of nodes. We show that lossless expansion implies polynomial time online matching and even polylogarithmic time ``approximate'' matchings.

\subsection*{Online matching game}

For the sake of illustration, we view left nodes as clients and right nodes as servers.
The bipartite relation models the fact that a client can only be satisfied by certain servers. 
If the graph has offline matching up to $K$ elements, then for every set of at most $K$ clients, one can assign unique servers. 
In {\em online matching } up to~$K$, clients arrive and may also depart, releasing the assigned match. 
At most $K$ clients may simultaneously need access to a server.  We also use a relaxed notion, in which a server may be assigned to up to~$\ell$ clients.
The formal definition uses a game.

\medskip
\noindent
\textbf{Game.}
The game with parameters $K$ and $\ell$ is played on a fixed graph. Two players, called Requester and Matcher, know this graph and alternate turns. 
Together they maintain a subset $M$ of edges, which is initially empty.
Requester starts. At his turn, he may  remove zero or more edges from~$M$.
After this, $M$ should contain at most $K-1$ edges.  Also, he must select a left node~$x$. 
At her turn, Matcher may add an edge to~$M$. 
After this, $x$~should be incident on an edge of~$M$, and each right node must be incident on at most $\ell$ edges from~$M$. 
If these conditions are not satisfied, then Matcher loses. 

\label{page-basic-game}
\begin{definition*}
  A graph has {\em online matching} up to~$K$ elements with load~$\ell$, 
  if Matcher has a strategy in the above game in which she never loses.
  If the load $\ell$ is omitted, then $\ell = 1$ is assumed.
\end{definition*}

\noindent
A fully connected bipartite graph with right size $K$ has online matching up to $K$. 
Both graphs below have offline matching up to~$2$ elements, but neither has online matching up to~$2$ elements\footnote{\label{foot:matchingGameExample}
  Requester wins the game on the right graph above by selecting the top and middle left nodes, releasing the top node, and requesting the bottom node. 
  Indeed, Matcher must answer the top node request with the top right node, or he loses immediately after the bottom node is requested. 
  But after the middle node is requested, the bottom node can not be matched anyways. 
}. 

\newcommand{\examplesMatchingGame}{
  \begin{center}
    \begin{tikzpicture}[scale=.8]
    \node[pt] (a) at (0,1)  {};
    \node[pt] (b) at (0,0)  {};
    \node[pt] (c) at (0,-1) {};
    \node[pt] (u) at (3,.5)  {};
    \node[pt] (v) at (3,-.5)  {};
    \draw (a) -- (u);
    \draw (b) -- (u);
    \draw (a) -- (u);
    \draw (b) -- (v);
    \draw (c) -- (v);
  \end{tikzpicture}
  \qquad
  \qquad
  \qquad
    \begin{tikzpicture}[scale=.8]
    \node[pt] (a) at (0,1)  {};
    \node[pt] (b) at (0,0)  {};
    \node[pt] (c) at (0,-1) {};
    \node[pt] (u) at (3,.5)  {};
    \node[pt] (v) at (3,-.5)  {};
    \draw (a) -- (u);
    \draw (b) -- (u);
    \draw (a) -- (u);
    \draw (a) -- (v);
    \draw (b) -- (v);
    \draw (c) -- (v);
  \end{tikzpicture}
  \end{center}
}
\examplesMatchingGame

\begin{remark*}
    The objective in the game is different from the extensively studied dynamic and online matching problems in the literature, 
    in which the graph is not fixed and the request consists of a left node with its edges (adversarially chosen). 
    In dynamic matching, matches may be revoked. 
    In both areas, the objective is to maintain a matching for as many active requests as possible, see for example the presentations in~\cite{hen-kha-pau-sch:c:dynamicmatching,williams:c:dynmatching,BehnezhadGhafari2024FOCS}.  
    Our definition is incomparable to this. 
    On one side, the definition is weaker because the graph is fixed and known to the players. 
    On the other side, it is stronger because we require a matching for all requested nodes and we may not change previously assigned matches. See~\cref{sec:relatedWork} for more related results. 
\end{remark*}

Feldman, Friedman, and Pippenger~\cite[proposition 1]{fel-fri-pip:j:networks}  have shown that if a graph has $2$-expansion up to~$2K$, then it has online matching  up to~$K$. 
By a similar argument, $1$-expansion up to~$K$ implies online matching  up to~$K$ with load~$3$, see~\cref{s:slowmatch}.
Unfortunately,  matches are computed in time exponential in~$K$.

We show that a graph with expansion factor equal to a large fraction of the left degree (i.e., a lossless expander) has  polynomial-time online matching and, moreover, if we allow small load, it has logarithmic time online matching.
\begin{proposition} \label{prop:poly_time_matching}
  If a graph with $N$ left elements and left degree $D$ has $(\tfrac 2 3 D + 2)$-expansion up to~$K$, then it has online matching up to~$K$ and each match of the strategy  is computed in time~$\poly(N)$. 
\end{proposition}

\begin{theorem} \label{th:DynamicMatching}
  If a graph with $N$ left elements and left degree $D$ has $(\tfrac 2 3 D + 2)$-expansion up to~$K$, then it has online matching up to~$K$ with load $O(\log N)$, 
  and the matching strategy requires $O(D \log N)$ amount of computation to compute each match and process each retraction.\footnote{
    The runtimes are expressed assuming the Word RAM model with cell size = 1 bit. 
    If the cell size is $\Theta(\log N)$, which is common in the graph algorithms literature, then the runtime for each match is $O(D)$.
  } 
\end{theorem}

The algorithms receive the game state in the most natural way, see the first paragraphs of section~\ref{sec:matchpolytime} and the definition in \cref{sec:fastMatch} below for details. 
The algorithms use a data structure to store information for faster computation of future matches.  The load $\ell = O(\log N)$ can be reduced to $1$, by making $\ell$ {\em clones} of the right side connected to the left side as in the original graph.  In this graph both the left degree and the right size increase by a factor of $\ell$, and the runtime remains $O(\text{(left-degree)}\cdot \log N)$.

Note that in~\cref{th:DynamicMatching}, the runtime for computing a match is double exponentially faster than in~\cite{fel-fri-pip:j:networks}.
The table below provides a summary of various ``Hall-type"  results.
If a graph with $N$ left nodes and left degree $D$ satisfies the expansion condition in column 1, then it has matching with the features in columns 2, 3, and~4.

\renewcommand{\arraystretch}{1.5}
\medskip

  \begin{center}
\begin{footnotesize}
\begin{tabular}{|l|l|l|l|l|}

\hline
\vspace{0.3mm}
expansion  up to~$K$ & matching up to~$K$  &  load & runtime per match & reference   \\
\hline \hline

$1$ & offline & $1$   &  N/A & Hall's Theorem\\ 
\hline
$1$ &  online  & $3$ & $N^{K+O(1)}$ & \cite{fel-fri-pip:j:networks}, \Cref{c:ffp}\\
\hline
$2D/3+2$ & online& $1$  & $\poly(N)$   &  \Cref{prop:poly_time_matching}\\
\hline
$2D/3+2$ & online & $O(\log N)$ & $O(D \log N)$ & \Cref{th:DynamicMatching} \\
 \hline
\end{tabular}
\end{footnotesize}
 \end{center}
\renewcommand{\arraystretch}{1}

These results add a new entry to the list of wonderful properties of lossless expanders (for an overview, see~\cite[Chapter 10]{hoo-lin-wig:j:expander} or~\cite[section 1.3]{cap-rei-vad-wig:c:conductors}). 

\bigskip
\noindent

In applications, parameters $N$ and $K$ are given, and a graph is needed with left size~$N$ and online matching up to $K$. It is desirable that the graph has:
\\-- few right nodes, (ideally, close to $K$),   
\\-- few edges, (ideally, close to $N$), and 
\\-- fast matching time (ideally, the algorithm should be ``local," i.e., it should inspect only  the neighborhoods of a few left nodes). 

We come close to this goal. 
There exist non-explicit constructions of lossless expanders (see \cref{lem:non-explicitexpander}) in which the right size is $O(K\log N)$ and the number of edges is $O(N\log N)$. 
Hence, the 3 objectives (the right size, number of edges, and matching time) are simultaneously optimal up to a $O(\log N)$ factor. 
Known explicit constructions of lossless expanders (see theorem~\ref{th:explicitexpander}) provide graphs with online matching in which the right size, number of edges, and matching time are optimal up to $\quasipoly(\log N)$. 

   Below, we present applications to communication networks. In the companion paper~\cite{companion-DS}, we present applications to set data structures (which rely on an easier game that allows a small number of revocations).

\subsection*{Proof ideas}

For theorem~\ref{th:DynamicMatching} (proven in section~\ref{sec:fastMatch}),  the idea  is to combine 2 matching strategies.

  The first one is a greedy strategy for an online matching game, called $T$-round matching,  which is the same as the basic game except that it has the constraint that it is played for $T$ rounds, where $T$ is an arbitrary but fixed number (known beforehand by both players).  We explain the basic idea using a simpler game, called \emph{incremental matching}.  In this game in which Requester only adds requests (no removals) and is played for $K$ rounds. The following greedy strategy from~\cite{mus-rom-she:j:muchnik} works on an $\lceil \log K \rceil$-clone of a graph with 1-expansion up to $K$. Let the clones be numbered. When a left node is requested, Matcher replies with an edge to any available neighbor in the clone with the smallest index.  
  We explain that this strategy works. Let $R$ be the set of requests that are matched outside the 1st clone. These matches have at least $\#R$ neighbors by 1-expansion, and all of them must have been assigned to other requests (for which the 1st clone was used). Thus the 1st clone allocates at least half of the requests. By the same argument, the 2nd clone allocates at least half of the remaining requests, etc. 
  Thus all requests are eventually matched using $\lceil \log K \rceil$ clones. Merging the clones, we obtain matching with load $\lceil \log K \rceil$.  With a variant of this argument, this extends from incremental matching to $T$-round matching game (which also allows matchings retractions), see the companion paper~\cite{companion-DS} for details. 

The second strategy is the one used to prove proposition~\ref{prop:poly_time_matching}. 
The idea is to assign arbitrary free neighbors for all requested nodes, with the exception of left nodes that have at least 1/3 of their neighbors already assigned to other nodes.  
  Such a critical node is ``protected'' by receiving a virtual match. (Which is used as the match when requested.) The large expansion implies that not too many critical nodes exist and the unique neighbor property of lossless expanders allow to assign virtual matches in polynomial time. 

Finally, the 2 strategies are combined, namely 
 the $\poly(N)$ time procedure from \cref{prop:poly_time_matching} and the greedy procedure for $T$-round matching. The latter runs in time $O(D\log N)$ as required, but can not assign matches for a few problematic left nodes. Fortunately, a small subset containing these nodes can be identified well in advance, i.e., many requests before such a problematic request might happen, and handled by the slower procedure on a separate copy of the graph. This leads to a small amortized runtime. We next interlace the two procedures on further copies of the graph. This allows de-amortization and gives worst-case runtime $O(D \log N)$.

\subsection*{Application: non-blocking connectors.}

 Switching networks are used for the transfer of information between a large number of nodes. 
For example, in the early days of telephone networks, when there were only a few phones in a town, people made pairwise connections between all phones. 
When the number of phones grew, this was no longer feasible, and switching stations were introduced. 
Their theoretical study was initiated by Shannon~\cite{sha:j:switchingnetworks} and was the motivation for introducing expander graphs~\cite{bas-pin:j:networks, mar:j:concentrators}. 
Currently there is a large literature, both in the engineering and the theoretical computer science fields. 
See the book~\cite{hwa:b:networks} for history and background. 

Switching networks are still used in engineering where many components need to communicate.
Unlike telephone networks, most applications concern the bipartite variant with inputs on one side and outputs on the other side, see~\cite[section~1.1]{hwa:b:networks}. 
In such graphs, the aim is to connect all output nodes to any permutation of input nodes using node disjoint paths.

An {\em $N$-network} is a directed acyclic graph in which $N$ nodes are called inputs and $N$ nodes are called outputs. Its size is the number of edges.
A {\em rearrangeable} $N$-network is such a network in which for every 1-to-1 function $f$ from outputs to inputs, there exist $N$ node disjoint paths that connect each output node $i$ to the input node $f(i)$.

For example, a fully connected bipartite graph with left and right sets of size $N$ defines a rearrangeable $N$-network with $N^2$ edges. Another example is given in figure~\ref{fig:gnetwork}.  
The goal is to construct networks with a minimal number of edges. 
Since there are $N!$ different mappings,
the minimum is at least $\log N!$, which is at least $N(\log N - 2)$ by Stirling's formula.

In many applications, the connection problem needs to be solved dynamically. 
For this, 2 closely related requirements exist, which are called \emph{strict-sense non-blocking} and \emph{wide-sense non-blocking}.\footnote{
  For strict-sense non-blocking, a path between a pair (output, input) making a request can be found given any paths between previous pairs; in wide-sense, a path satisfying a new request is guaranteed only if the previous paths were established by a specific algorithm, see~\cite{hwa:b:networks} and our definition via a game.
  The Clos network and the connector from \cite{aro-lei-mag:j:network} mentioned below satisfy the stronger property.
}
We follow the~\cite{fel-fri-pip:j:networks} paper which uses the wide-sense variant, which is weaker, and which we formally define below by a game. On the other hand, the paper uses generalized connectivity, which allows 1-to-many (instead of 1-to-1) connectivity: several output nodes may be connected to the same input, but each output is connected to at most 1 input. 
In terms of broadcasting, this means that several outputs can listen to the same input.

\medskip
\noindent
\textbf{Connection game.}
The game is played on an $N$-network. Two players, called Requester and Connector, both know the network and  alternate turns. They maintain a set of at most $N$ trees. 
The root of each tree must be an input and the leaves must be outputs. The trees must be node disjoint.
Initially, the set of trees is empty. 
 Requester starts. 

At his turn, Requester may remove zero or more trees.
Afterwards, he may select an input $x$ and an output $y$ such that $y$ does not lay on any of the trees. 

At her turn, Connector may create or extend a tree so that the above conditions are satisfied. Afterwards, there should be a tree in which $x$ is the root and $y$ is a leaf.
If this is not true, she loses.

\begin{definition*}
  A {\em wide-sense non-blocking generalized $N$-connector}  is an $N$-network in which Connector has a strategy in which she never loses.
  We refer to such a network as {\em $N$-connector}.
\end{definition*}

\begin{figure}
\begin{center}
 \begin{tikzpicture}[scale=.8]
  \node[pt] (a) at (0,2)  {};
  \node[pt] (b) at (0,0)  {};
  \node[pt] (c) at (0,-2) {};
  \node[pt] (a1) at (2,2.5)  {};
  \node[pt] (a2) at (2,1.5)  {};
  \node[pt] (b1) at (2,0.5)  {};
  \node[pt] (b2) at (2,-0.5)  {};
  \node[pt] (c1) at (2,-1.5)  {};
  \node[pt] (c2) at (2,-2.5)  {};
  \node[pt] (ra1) at (3,2.5)  {};
  \node[pt] (ra2) at (3,1.5)  {};
  \node[pt] (rb1) at (3,0.5)  {};
  \node[pt] (rb2) at (3,-0.5)  {};
  \node[pt] (rc1) at (3,-1.5)  {};
  \node[pt] (rc2) at (3,-2.5)  {};
\node[pt] (ra) at (5,2)  {};
  \node[pt] (rb) at (5,0)  {};
  \node[pt] (rc) at (5,-2) {};

  \draw (a) -- (a1);
  \draw (a) -- (b1);
  \draw (b) -- (a2);
  \draw (b) -- (c1);
  \draw (c) -- (b2);
  \draw (c) -- (c2);
   \draw (a1) -- (ra1);
    \draw (a1) -- (ra2);
 \draw (a2) -- (ra1);
 \draw (a2) -- (ra2);
\draw (b1) -- (rb1);
    \draw (b1) -- (rb2);
 \draw (b2) -- (rb1);
 \draw (b2) -- (rb2);
 \draw (c1) -- (rc1);
    \draw (c1) -- (rc2);
 \draw (c2) -- (rc1);
 \draw (c2) -- (rc2);
\draw (ra) -- (ra1);
  \draw (ra) -- (rb1);
  \draw (rb) -- (ra2);
  \draw (rb) -- (rc1);
  \draw (rc) -- (rb2);
  \draw (rc) -- (rc2);

\end{tikzpicture}
  {\small
  \caption{
    A connector with 3 inputs (left) and 3 outputs (right), depth 3, and size 24. 
  }}
\end{center}
\label{fig:gnetwork}
\end{figure}

A fully connected bipartite graph is an $N$-connector.
An $N$-connector was given in~\cite{fel-fri-pip:j:networks} that has $O(N \log N)$ edges. 
This is optimal within a constant factor.  The graph is explicit but the path finding algorithm (which is the algorithm that computes  Connector's reply) is very slow. 
Afterwards, in~\cite{aro-lei-mag:j:network} an explicit $N$-connector is constructed of size $O(N \log N)$ 
in which also the runtime of the path finding algorithm is $O(\log N)$, and this is optimal within a constant factor. 
See~\cite{aro-lei-mag:j:network} or \cite[chapter 2]{hwa:b:networks}.

The {\em depth} of a network is the length of the longest path between an input and an output. The $N$-connectors from~\cite{fel-fri-pip:j:networks} and~\cite{aro-lei-mag:j:network} have depth $\Theta(\log N)$.
Our results are about  constant depth $N$-connectors. 
In~\cite{pip-yao:j:connector} it is shown that $N$-connectors of depth $t$ have at least $tN^{1+1/t}$ edges. 
In~\cite{fel-fri-pip:j:networks} non-explicit constructions of $N$-connectors are given of size $O(N^{1+1/t} \log^{1-1/t} N)$, but again the path finding algorithm runs in time exponential in~$N$. 
They ask whether a generalized connector exists with small size and an efficient path finding algorithm. They do not specify explicitly, but ``small size" is usually considered to be a value that is $N^{1+1/t} \cdot N^{o(1)}$, see~\cite{wig-zuc:j:expander}, and ``efficient" should ideally mean that the runtime is $\poly (\log N)$. 
Some explicit constant-depth $N$-connectors are known with path finding algorithms running in time $\poly(\log N)$, 
but their size is not optimal, see~\cite[chapter 2]{hwa:b:networks}. 
For instance, for odd $s$, the optimal $s$-stage Clos network of minimal size  has depth $t=2s-1$ and size $\Theta_t(N^{1+4/(t+3)})$.

In~\cite[Th. 5.4]{wig-zuc:j:expander} an explicit construction of size $N^{1+1/t} \exp((\log  \log N)^{O(1)})$ was obtained, 
but the path finding algorithm is the same slow one from~\cite{fel-fri-pip:j:networks}.



In section~\ref{sec:connectors}, we present a non-explicit constant depth $N$-connector whose size is optimal up to factors $\poly(\log N)$ and with a path finding algorithm running in time $\poly(\log N)$. Here we assume that the input of this algorithm is a description of the state of the connection game that includes the graph\footnote{For each node the input specifies the degree and a list of neighbors in arbitrary order.} and the algorithm may use a data structure.

\begin{corollary}\label{cor:nonexplicitConnector}
  For all $t$ and $N$, there exists an $N$-connector of depth $t$ 
  and size 
  \[
    O(N^{1+1/t} \log^3 N)
  \]
  with a $\poly(\log N)$ time path finding algorithm.
\end{corollary}

Unfortunately, the implicit constants in the big-O notation are large and the construction seems not (yet) practical. 
Even for a variant with $\poly(N)$ time path finding (instead of $\poly(\log N)$), we obtain size $\frac {2^9} {t} \log^2 N \cdot N^{1+1/t}$, 
for $N \ge 2^{7t + 2}$.

An $N$-connector is {\em explicit} if the $i$-th neighbor of a node is computed in time $\poly(\log N)$. 
We present an explicit connector with small size and  a path finding algorithm running  in $\quasipoly( \log N)$  time.
\begin{corollary}\label{cor:explicitConnector}
  For all $t$ and $N$, there exists an explicit $N$-connector of depth $t$, size 
  \[
    N^{1+1/t} \exp(O(\log^2 \log N)),
\]  
  with a path finding algorithm with runtime $\exp(O(\log^2 \log N))$. 
\end{corollary}

{\bf Proof idea.} The construction is the same as in Feldman, Friedman and Pippenger~\cite{fel-fri-pip:j:networks} (and also used in \cite[Th. 5.4]{wig-zuc:j:expander}): it consists of layers obtained by composing  bipartite graphs that have sufficient expansion to allow online matching, see~\cref{fig:connector_construction} for the idea. The double exponential faster time for path finding is due to the faster online matching strategy of the graphs in theorem~\ref{th:DynamicMatching}. The improved size (compared to \cite[Th. 5.4]{wig-zuc:j:expander}) of the explicit construction in~\cref{cor:explicitConnector} is due to the better lossless expanders from~\cite{guv:j:extractor}.

\subsection*{Open questions}

Theorem~\ref{th:DynamicMatching} assumes large expansion, logarithmic load, and that the algorithm uses a data structure. 
We do not know if we can strengthen any of these assumptions. The strongest claim that we can not refute is the following. 

\medskip
\noindent
\textbf{Open question~1.} Does there exist $e$ and $\ell$ with $e+\ell \le 3$ 
such that each graph with $e$-expansion up to $K$ has online matching up to $K$ with load~$\ell$, 
where the time for computing matches is $O(D\log N)$ without using a data structure.\footnote{
  Recall that the input is the state of a game with the underlying graph. 
  We do not need to process retractions, since there is no data structure.
}

\medskip
The following open question is weaker than the above open question. However, if true, then the size of explicit connectors with depth~$t$  can be further improved
to $N^{1+1/t}\poly(\log N)$ instead of $N^{1+1/t}\quasipoly(\log N)$. (Recall that the lower bound is $tN^{1+1/t}$.) 

\medskip
\noindent
\textbf{Open question~2.}
Does $1$-expansion up to $K$ imply online matching up to $K$ with load $O(\log N)$ in which matches and retractions are processed in time $\poly(D\log N)$ with a data structure?

\medskip
\noindent
The above claim regarding connectors follows from the explicit graph constructed in the companion paper~\cite{companion-DS}, 
with 1-expansion up to $K$, degree $\widetilde O(\log^2 N)$, and right size $(1+\eps)K$.
In contrast, the {state-of-the-art explicit graph with $(2D/3 + 2)$-expansion up to K  has  $D = \poly(\log N) \cdot 2^{O((\log \log K)^2)}$ and $\#R = K \cdot \poly(D, \log N)$ (see~\cref{th:explicitexpander}, obtained from~\cite[theorem 18]{lu-oli-zim:c:optimalcode}).

A further relaxation of the last open question is to require $\poly(N)$ runtime instead of $\poly(\log N)$. 
Proving that such an algorithm does not exist requires some hardness assumption because if P = NP, the algorithm  from~\cite[proposition 1]{fel-fri-pip:j:networks} 
 runs in time $\poly(DN)$.

\subsection*{Road map}

\begin{center}
  \begin{tikzpicture}[xscale=2, every node/.style={draw,rectangle, align=left}]
    \node (Texp) at (2, 2.5) {Prop.~\ref{cor:TroundMatching} \\ T-round matching\\ polylog time \\ companion paper~\cite{companion-DS}};
    \node (prop11) at (-1, 2.5) {Prop.~\ref{prop:poly_time_matching} \\ online matching \\ poly time \\section~\ref{sec:matchpolytime}};
    \node (thm12) at (-1, 0) {Thm.~\ref{th:DynamicMatching}, \\ online matching\\ polylog time \\section~\ref{sec:fastMatch}};
    \node (appConn) at (-1, -2.25) {Application \\connectors \\ section~\ref{sec:connectors}};
    \draw[->,thick] (Texp) -- (thm12);
    \draw[->,thick] (prop11) -- (thm12);
    \draw[->,thick] (thm12) -- (appConn);
  \end{tikzpicture}
  \begin{center}
    Dependencies between results and applications.
  \end{center}
  
\end{center}

\section{Polynomial time online matching}\label{sec:matchpolytime}

We prove proposition~\ref{prop:poly_time_matching}. 
The algorithm is used in the faster $O(D\log N)$-time algorithm in~\cref{th:DynamicMatching}.

For notational simplicity, we prove it with a slightly stronger assumption: we assume expansion up to $K+1$ instead of~$K$. 
The original proposition is proven  similarly.\footnote{
  Define nodes to be critical if they have at least $D/3 + 1$ matched neighbors (instead of $D/3$), and follow the proof with extra $+1$'s and $-1$'s where needed. 
  In the second lemma below, bound the number $\#C$ of critical nodes by $K-1$ instead of $K$. 
  }
We state this variant. 

\begin{proposition} 
  If a graph with left size $N$ and left degree $D$ has $(\tfrac 2 3 D + 2)$-expansion up to~$K+1$
  then it has an online matching algorithm up to~$K$ in which each match is computed in time~$\poly(N)$. 
\end{proposition}

\noindent
Recall the online matching game with $\ell=1$. 
Requester and Matcher maintain a set $M$ of edges. They alternate turns, and at their turn they do the following.
\begin{itemize}
  \item 
     Requester removes edges from $M$ so that~$\#M \le K-1$. He also selects a left node~$x$. We say that he {\em requests}~$x$. 
   \item 
     If $x$ is not covered by $M$, then Matcher must reply by adding an edge $(x,y)$ to~$M$. After this, $M$ must be a matching. 
     The right node $y$ is the {\em match} of~$x$. 
\end{itemize}
The aim of Matcher is to provide correct replies indefinitely.

A matching algorithm has as input the state of the game after Requester's move, Requester's move itself, and some data structure (which stores information to speed up computations in future rounds). 
The state of the game consists of $N$, $K$, the graph, and the matching.

\bigskip
\noindent
The proof uses 2 technical lemmas. 
Let $\N(S)$ be the set of neighbors of a set~$S$ of nodes.
Given a set $S$ of left nodes, we call a right node $y$ a {\em unique} neighbor of $S$ if $y$ has precisely 1 left neighbor in~$S$. 
The following lemma holds in any bipartite graph with left degree~$D$. 

\begin{lemma}\label{lem:privateNeighbors}
  The number of unique neighbors of $S$ is at least $2\#\N(S) - D\#S$. 
\end{lemma}

\begin{proof}
  We need to lower bound the number $u$ of unique neighbors. 
  The number of vertices in $\N(S)$ that are not unique, equals $\#\N(S) - u$. 
  There are $D\#S$ edges with an endpoint in~$S$. 
  For each such edge, the right endpoint is either unique or has at least 2 neighbors in $S$. 
  Hence,
  \[
    D\#S \ge u + 2(\#\N(S) - u).
  \]
  The lower bound of the lemma follows by rearranging.
\end{proof}

\noindent
The following lemma holds for graphs satisfying the assumption in the proposition. 

\begin{lemma}
  Let $Y$ be a subset of right nodes with $\#Y \le 2K+1$. If a left set $C$ contains only nodes $x$ with $\# \N(x) \cap Y \ge D/3$, then $\#C \le K$. 
\end{lemma}

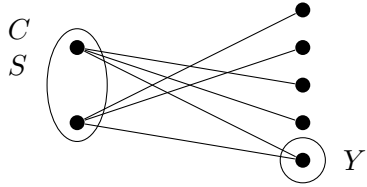
\begin{center}
  \begin{tikzpicture}[yscale=0.5]
    \node[pt] (u) at (0,3) {};
    \node[pt] (v) at (0,1) {};
    \foreach \ind in {0,...,4}{
      \node[pt] (r\ind) at (3,\ind) {};
    }
    \draw (v) -- (r4);
    \draw (v) -- (r3);
    \draw (u) -- (r2);
    \draw (u) -- (r1);
    \draw (u) -- (r0);
    \draw (v) -- (r0);
    \draw (0,2) ellipse (0.4cm and 1.5cm);
    \node[anchor=east, text width=1em] at (-.4,3) {$C$\\$S$};
    \draw (3,0) ellipse (0.3cm and 0.6cm);
    \node[anchor=west] at (3.4,0) {$Y$};
  \end{tikzpicture}
\end{center}

\begin{proof}
  Suppose $C$ contains at least $K+1$ elements, and let $S$ be a subset of $C$ of size exactly~$K+1$. 
  By assumption on $C$, each of its nodes has at most $\tfrac 2 3 D$ neighbors outside~$Y$. 
  Thus, by expansion,
  \[
    (\tfrac 2 3 D + 2)\# S \le \#N(S) \le \# Y + \tfrac 2 3 D \# S.
  \]
  This simplifies to $2\#S \le \#Y$. But this contradicts $\#S = K+1$ and $\#Y \le 2K + 1$. 
\end{proof}

\begin{proof}[Proof of the proposition.] 
  The idea of the matching algorithm is to assign a ``virtual match'' to left nodes for which at least $D/3$ neighbors are matched. 
  Note that there are 2 types of matches to which we refer as standard and virtual matches. In the $D/3$ bound, we count both types. 
  No other nodes can not be matched to a virtual match of a node. Virtual matches are stored in the data structure.

  Left nodes with at least $D/3$ matched neighbors (of both types) are called {\em critical}. 
  A virtual match will be assigned to a left node $x$ if and only if $x$ is critical and has no  match. 

  Requester's move consists of a list of retracted edges and the requested left node.
  For each retracted edge, a retraction algorithm is ran. Afterwards, 
  the match generation is done. 

  \medskip
  \noindent
  {\em Algorithm for retracting a match $(x,y)$.}
  If $x$ is critical, then declare $y$ to be a virtual match. 
  Otherwise, retract the match and retract all virtual matches of left nodes with less than~$D/3$ matches. 
 
  \begin{center}
    \qquad\qquad
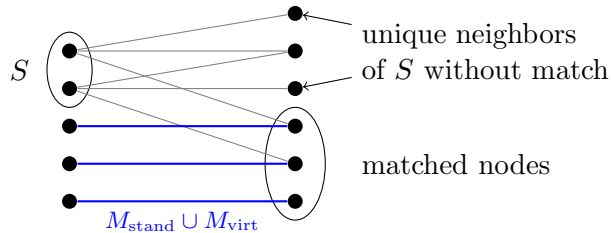
    \begin{tikzpicture}[yscale=0.5]
      \foreach \ind in {0,...,5}{
	\node[pt] (r\ind) at (3,\ind) {};
      }
      \foreach \ind in {0,...,4}{
	\node[pt] (l\ind) at (0,\ind) {};
      }

      \draw (0,3.5) ellipse (0.3cm and 1cm);
      \node[anchor=east] at (-.4,3.5) {$S$};
      \draw (3,1) ellipse (0.4cm and 1.5cm);
      \node[anchor=west] at (3.75,1) {matched nodes};

      \foreach \aa/\bb in {4/5, 4/4, 4/2, 3/4, 3/3, 3/1}{
	\draw[gray] (l\aa) -- (r\bb);
      }
      \foreach \ind in {0,1,2}{
	\draw[blue,thick] (l\ind) -- (r\ind);
      }
      \node[anchor=north,blue] at (1.5,0) {\footnotesize $M_\textnormal{stand} \cup M_\textnormal{virt}$};
      \node[text width=3.3cm, anchor=west] (un) at (3.75,4) {unique neighbors of~$S$ without match};
      \draw[->] (un) -- (r5);
      \draw[->] (un) -- (r3);
    \end{tikzpicture}
    \\Virtual matches of critical nodes are unique neighbors.
  \end{center}

  \bigskip
  \noindent
  {\em Generating a matching for a request~$x$.}
  If the request is a critical node, then its virtual match $y$ is returned, and thus $y$ is now a standard match.  
  Otherwise, $x$ is matched to any neighbor that does not have a match (of either type). 
  (Such a neighbor exists because a non-critical node has more than $D-D/3$ unmatched neighbors.)

  After this, there might be critical nodes which do not have a match.
    Let $S$ be the set of such nodes.  
  Virtual matches for these elements are assigned 1 by 1 as follows. 

  Select an unmatched right node $y$ that has exactly 1 neighbor in $S$. 
  Below we explain that such a $y$ always exists. 
  Let $x$ be this single neighbor.
  Remove~$x$ from~$S$, and declare $y$ to be its virtual match. 
  Add to $S$ all new critical nodes without a match. 
  Keep repeating until $S$ is empty.
  (This must happen, because an element can be added to~$S$ at most once.)
  This finishes the description of the matching algorithm.

  \bigskip
  \noindent
  Note that the 2 algorithms above require $\poly(DN)$ amount of computation. 
  We may assume that $D\le N$, since otherwise the proposition is trivial. Hence, the runtime is $\poly(N)$.   
  In the presentation of the algorithm a claim was made: the set $S$ of critical nodes always has at least 1 unique and unmatched neighbor.  
  If this is true, the online matching algorithm always produces matches and the proposition is proven.  

  \bigskip
  \noindent
  We first prove 2 other claims. 

  {\em Proof that at any moment, at most $K$ nodes are critical.} 
  In the above algorithm, matches are added 1 by~1. 
  Assume that just before allocating a match there are at most~$K$ critical nodes. 
  Then the number of standard and virtual matches is at most $K + K$ (and in fact, it is 1 less). 
  Let $Y$ be the set of matched right nodes with the new match included, thus $\#Y \le 2K+1$. 
  By the second lemma, there are still at most~$K$ critical nodes.

  {\em Proof that at any moment, all nodes in~$S$ have exactly $\lceil \tfrac D 3 \rceil$ matched neighbors.} 
  By construction a node is placed in~$S$ when it has at least $\tfrac D 3$ neighbors. 
  This condition is checked each time after a match is assigned, thus when a node is added to $S$, it has exactly $\lceil \tfrac D 3\rceil$ neighbors. 
  As long as $S$ is nonempty, a virtual match~$y$ is given to a left node $x$ such that $y$ has no other neighbors in~$S$, and then $x$ is removed. 
  Thus for all other nodes in~$S$, the number of matched neighbors remains the same.
 
  {\em Proof that in the above matching algorithm, an unmatched node $y$ exists that has exactly~1 left neighbor in~$S$.} 
  Since all nodes in $S$ are critical, we have $\# S \le K$. 
  By the assumption on expansion, $\#\N(S) \ge (\tfrac 2 3 D + 2) \#S$. 
  By the first lemma, $S$ has at least $(\tfrac 1 3 D + 4)\#S$ unique neighbors. 
  At most $\lceil \tfrac 1 3 D\rceil\#S$ of the unique neighbors can be matched, by the previous point. 
  Hence, at least $3\#S$ right nodes are unique and unmatched. 
  Thus, if $\#S \ge 1$, the required right node $y$ exists, and if $\#S = 0$, no unique neighbor is needed.
  This finishes the proof of the claim inside the algorithm, and hence, the proposition is proven.
\end{proof}

\section{Fast online matching}\label{sec:fastMatch}

We prove the main result, \cref{th:DynamicMatching}. The proof relies  
on a proposition from the companion paper~\cite{companion-DS} about $T$-round matching, 
which is defined by the same game, except that Matcher already wins if he can reply $T$ times (instead of indefinitely). 
Graphs are given by adjacency lists, and checking whether an edge is in the matching happens in $O(1)$ time.

\begin{definition*}
  We call a matching strategy {\em fast} if the strategy can be presented by a retraction and a match generation algorithm as explained in the previous section 
  and the runtime of both algorithms is $O(D \log N)$, where $N$ and $D$ are the left cardinality and degree. 
\end{definition*}

\begin{proposition}[\cite{companion-DS}]\label{cor:TroundMatching}
  If a graph has $1$-expansion up to~$K$, then it has fast $T$-round matching up to~$K$ with load $O(\log T)$.
\end{proposition}

The matching strategy of \cref{th:DynamicMatching} combines an $O(D\log N)$ time greedy strategy from~\cref{cor:TroundMatching} with the $\poly(N)$ time strategy of section~\ref{sec:matchpolytime}.
The greedy strategy allocates most matches, while the polynomial one is used for a few problematic requests that are anticipated well in advance.
 
Recall that in fast online matching we use a data structure to compute matches. 
We consider a relaxed notion of fast matching that besides algorithms to generate matches and process retractions, 
also has a {\em preparation} algorithm. This algorithm is run at regular intervals and does not need to be fast. 
This algorithm prepares the data structure for fast computation of future matches. 

\begin{definition*}\label{def:}
  We say that a graph with left size $N$ and left degree $D$ has fast online matching with {\em $T$-step preparation} if there exists an online matching algorithm that 
  computes matches and processes retractions in time $O(D\log N)$ with the help of a preparation algorithm. This algorithm runs in time $O(T)$ and is ran 
  each time after $T$ matches have been assigned. 
\end{definition*}

\begin{remark*} 
A graph with fast online matching with preparation and load $\ell$ has fast online matching in the amortized sense. De-amortization with load $2\ell$ is obtained as follows. A 2-clone of such a graph has fast online matching (in the standard worst-case) because blocks of $T$-subsequent requests can alternatingly be given to the copies: 
\begin{center}
  \tikzstyle{li}=[blue, line width=2]
  \begin{tikzpicture}[yscale=.9,xscale=.75]
    \node[anchor=east] at (-1.6,0) {1st copy};
    \node[anchor=east] at (-1.6,-1) {2nd copy};
    \node [anchor=south] at (2, 0.1) {preparation};
    \node [anchor=south] at (2, -.9) {$T$ matches};
    \foreach \y in {0, -1}{
      \draw[dashed] (-1.5,\y) -- (-0.8,\y);
      \draw[dashed] (13.1,\y) -- ( 13.8, \y);
    }
    \draw[li] (0,0) -- (4,0);
    \foreach \coord in {-.5, -.2, 4.2, 4.5, ..., 8.4, 12.5, 12.8}{
	    \node[pt]  at (\coord, 0) {};
	  }
    \draw[li] (8.3,0) -- (12.3,0);
    \draw[li] (-.6,-1) -- (-.1,-1); 
    \foreach \coord in {0.1, 0.4, ..., 4.2, 8.4, 8.7, ..., 12.6}{
	    \node[pt]  at (\coord, -1) {};
	  }
    \draw[li] (4.2,-1) -- (8.2,-1); 
    \draw[li] (12.5,-1) -- (12.9,-1); 
  \end{tikzpicture}
\end{center}
while one copy is used for assigning matches, the other can run its preparation algorithm (in little chunks at each request). Merging the clones doubles the load.
\end{remark*}

\begin{lemma}\label{lem:merging_fast_and_slow}
  Consider two graphs with the same left set of size $N$.
  If the first has $(\tfrac 1 2 D + 2)$-expansion up to~$K$ and the second has polynomial time online matching up to~$2K$, 
  then their union has fast online matching up to~$K$ with load $O(\log N)$ and $\poly(N)$-step preparation. 
\end{lemma}

\noindent
Before proving the lemma, we show that it implies the main result. 
This is not so hard to prove, because with a constant number of clones of the graph from the theorem, we obtain the graphs satisfying the conditions of the lemma. Here are the details.

\begin{proof}[Proof of \cref{th:DynamicMatching}.]
  The graph $G$ in the assumption of the theorem has degree $D$ and expansion $\tfrac 2 3 D + 2 \ge \tfrac 1 2 D + 2$.
  By proposition \ref{prop:poly_time_matching}, graph $G$ has polynomial-time online matching up to~$K$. 
  By applying the lemma to $G \cup G = G$, this graph has online matching up to $K/2$ with load $O(\log N)$ and $\poly(N)$ preparation time. 

  By the remark above on de-amortization, a 2-clone of $G$ has online matching up to $K/2$ with load $O(\log N)$. 
  Hence, a 4-clone of $G$ has such matching up to~$K$. 

  Therefore, the original graph of the theorem has matching up to $K$ with load $O(\log N)$,
  by multiplying by 
  $4$ the constant hidden in $O(\cdot)$. 
  The theorem is proven, except for the lemma.
\end{proof}

\begin{proof}[Proof of the lemma.]
  Let $F$ be the graph with $(\tfrac 1 2 D + 2)$-expansion   and let $G$ be the graph with polynomial time online matching. 
  We may assume that their right nodes are disjoint, because this affects the load by at most a factor 2.  

  The idea is to run the fast matching algorithm of proposition~\ref{cor:TroundMatching} on $F$ and use~$G$ as a safety buffer. 
  During the preparation, $G$ contains precomputed matches for nodes that are ``at-risk'' in the sense that they have many busy neighbors in~$F$. 
  The preparation and retraction algorithms share a set containing matches in~$G$, called the {\em retraction queue}.
  Let $T$ be a polynomial of $N$ that we determine later.
  
  \bigskip
  \noindent
  {\em Match generation for the first $T$ requests.}
  Apply the fast matching algorithm from \cref{cor:TroundMatching} using graph~$F$. 
  Since $F$ has $1$-expansion, we obtain matching with load~$O(\log T)$.
  
  \medskip
  \noindent
  {\em Preparation algorithm.} Run $G$'s retraction algorithm for all matches from the retraction queue. 
  Also run it for all precomputed matches from the previous preparation run that are not in the current matching.
  
  Let $A$ be the set of left nodes with at least $D/2$ matched neighbors (the at-risk nodes). 
  Compute the induced subgraph  $F'$ of $F$ containing the left nodes \emph{not} in $A$ and the right nodes with load~$0$.
  The set $A$ and graph $F'$ are fixed until the next preparation run (and are used for match generation). 

  Precompute matches in $G$ for all nodes in $A$. 
  Do this by generating requests 1-by-1 in any order. 
  (We soon explain that $G$'s matching algorithm will indeed produce matches.)

  \medskip
  \noindent
  {\em Match generation for request~$x$.} 
  If $x$ is in $A$, return its precomputed match in $G$. 
  Otherwise, run the fast $T$-round algorithm from \cref{cor:TroundMatching} on the graph $F'$. 
  (We soon explain that $F'$ has $1$-expansion up to~$K$.) 

  \medskip
  \noindent
  {\em Retracting $(x,y)$.} 
  If $(x,y)$ is in $G$, then add the edge to the queue.  Otherwise, run the retraction algorithm of~$F'$. 

  \medskip
  \noindent
  The value of $T$ is chosen to be a polynomial in $N$ large enough so that the preparation algorithm can be performed in time $T$. 
  By \cref{cor:TroundMatching}, the runtime of computing a match satisfies the conditions. 
  
  Above, 2 claims were made that need a proof. After this,  the lemma is proven, 
  because by construction, the load of all nodes in $G$ is bounded by $1$, and for $F$ it is bounded by $O(\log T)$. 

  \medskip
  \noindent
  \textbf{Proof that $F'$ has 1-expansion up to~$K$.}  Let $S$ be a set of left nodes in~$F'$ of size at most~$K$. 
  By expansion in~$F$, the set has at least $(\tfrac 1 2 D + 2)\#S$ neighbors in~$F$. 
  Since $A$ is removed, each element in $S$ has at most $\tfrac 1 2 D$ matched neighbors in~$F$. Thus, by expansion, the number of neighbors in $F'$ is at least
  \[
    (\tfrac 1 2 D + 2)\# S \,-\, \tfrac 1 2 D \# S \ge  \#S.
  \]

  \medskip
  \noindent
  \textbf{Proof that the polynomial-time matching algorithm precomputes matches for all the nodes in $A$.} 
  For this, we need to show that before each request, the size of $G$'s matching is at most $2K-1$.
  First we show that $\# A <  K$. Suppose that $\#A \ge K$ and let~$S$ be a subset of $A$ with exactly $K$ elements.
  Let $M$ be the set of matches in $F$.   Since all matches in $F$ are actual (not pre-computed), we have $\# M \le K$.
  Each node in $S$ has at most $\tfrac 1 2 D$ neighbors that are not covered by $M$. 
  Hence, the number of neighbors of $S$ in $F$ is at most
  \[
    \#\N(S) \le \# M + (\tfrac 1 2 D) \# S \le K  + (\tfrac 1 2 D) \# S.
  \]
  By the expansion of $S$ in $F$, we conclude that
  \[
    (\tfrac 1 2 D + 2) \# S \le \# \N(S) \le K  + \tfrac 1 2 D \# S. 
  \] 
  Hence, $2\#S \le K$, but this contradicts~$\#S = K$.
  Therefore, the assumption $\#A \ge K$ must be false. 
  
  The claim follows because less than $2K$ precomputed matches can exist simultaneously. 
  Indeed, there are less than $K$ matches computed in the current preparation run and also there are at most $K$ matches from previous runs (that became actual matches and have not been retracted).
\end{proof}

\section{Constant-depth connectors with fast path finding algorithms}\label{sec:connectors}

Graphs with online matching up to~$K$ can be composed into $N$-connectors of constant depth~$t$.
The following construction was given in \cite[Proposition 3.2]{fel-fri-pip:j:networks},
and obtains an almost minimal number of edges.

\begin{proposition}\label{prop:matchig2connector}
  Let $N$ be a $t$-th power of an integer, and let $C,D$ be constants. 
  Assume that for all integers $c < t$, there exist graphs with 
  $C N^{(c+1)/t}$ left nodes, at most $C N^{c/t}$ right nodes, left degree at most $D$, and online matching up to~$N^{c/t}$.
  Then, there exists an $N$-connector of depth $t$ with at most $tCD N^{1+1/t}$ edges.
\end{proposition}

\noindent
Recall that each connector has at least $tN^{1+1/t}$ edges (see \cite[proposition 2.1]{pip-yao:j:connector}). 
Hence, the above result is minimal within a factor $CD$.

\begin{remark*}
  To compute a path or extend a tree in this construction, at most $t$ matches need to be computed.
  Thus, for matching obtained from \cref{th:DynamicMatching} we obtain path finding in time $O(tD\log N)$. 
\end{remark*}

\noindent
For the sake of completeness, we present the construction and prove the proposition. 
First some observations about the static case are given.

An {\em $(N,N')$-network} is a directed acyclic graph with $N$ input and $N'$ output nodes.
A network is {\em rearrangeable} if every 1-to-1 mapping from outputs to inputs can be realized using node disjoint paths.

The following 2 lemmas are directly obtained from the definitions.

\begin{lemma}
  Consider a graph with $N$ left and $N'$ right nodes that has offline matching up to~$K$.
  The concatenation of this graph with a rearrangeable $(N', K)$-network is a rearrangeable $(N,K)$-network.
\end{lemma}

\begin{lemma}
  Consider $B$ rearrangeable $(N,N')$-networks with the same set of inputs and disjoint outputs.
  The union of these $B$ networks is a rearrangeable $(N, BN')$-network.
\end{lemma}

\newcommand{\drawChain}[1]{
     \foreach \i in {-#1,...,#1}{
       \node at (0, \i) {};
     }
 }
\newcommand{\drawThreeMatchingGraphs}[1]{
  \begin{scope}[xshift=10cm]
    \foreach \yyshift in {-2*#1cm-3cm, 0, 2*#1cm + 3cm}{
    \begin{scope}[yshift=\yyshift]
      \drawChain{#1}
    \end{scope}
  }
  \end{scope}
  \draw[fill=green!30,opacity=0.40]  (1,-3*#1-1.25) -- (1,+3*#1+1.25) -- (9,+3*#1+3.25) -- (9, #1+2.75) -- cycle;
  \draw[fill=blue!30,opacity=0.40]   (1,-3*#1-1.25) -- (1,+3*#1+1.25) -- (9,+1*#1+0.25) -- (9, -#1-0.25) -- cycle;
  \draw[fill=orange!30,opacity=0.40] (1,-3*#1-1.25) -- (1,+3*#1+1.25) -- (9,-#1-2.75)   -- (9, -3*#1-3.25) -- cycle;
}

\newcommand{\drawboxes}[1]{ 
  \begin{scope}[xshift=10cm]
    \foreach \yyshift in {-2*#1cm-3cm, 0, 2*#1cm + 3cm}{
      \begin{scope}[yshift=\yyshift]
	\draw[fill=gray!30,opacity=0.40]  (1,-#1-0.25) -- (1,#1+0.25) -- (9,#1-0.25) -- (9, -#1+.25) -- cycle;
	\begin{scope}[xshift=10cm]
	   \drawChain{1}
	\end{scope}
      \end{scope}
    }
  \end{scope}
}

\begin{figure}
  \centering
  \tikzstyle{pt}=[circle, fill=black, inner sep=1pt]
  \begin{tikzpicture}[every node/.style=pt, yscale=0.23, xscale=0.28]
    \begin{scope}[yscale=1.18]
    \drawChain{13}
    \end{scope}
    \begin{scope}[yscale=1.05]
    \drawThreeMatchingGraphs{4.5}
    \end{scope}
    \foreach \yshift in {-13, 0, 13}{
      \begin{scope}[xshift=10cm, yshift=\yshift cm, yscale=0.8]
         \drawThreeMatchingGraphs{1.5}
         \drawboxes{1.5}
      \end{scope}
   }
 \end{tikzpicture}
 {\small 
 \caption{\label{fig:connector_construction}
   Construction of a $27$-connector. Left column: 3 copies of a graph with online matching up to $9$. 
   Middle: 9 copies of a graph with online matching up to $3$. Right: 9 fully connected graphs. 
 }}
\end{figure}

\begin{proof}[Proof of~\cref{prop:matchig2connector}.] 
  Let $N = B^t$ for an integer $B$. The construction is illustrated for $B = t = 3$ in figure~\ref{fig:connector_construction}.

  \medskip
  \noindent
  {\em Construction of a rearrangeable $(C N, N)$-network.} 
  For every $c \le t$, we construct a $(C B^c, B^c)$-rearrangeable network of depth $c$ recursively.
  For $c = 1$, we use  the complete bipartite graph with left size $CB$ and right size $B$. 

  Suppose for some $c \ge 1$, we already have such a network $H$.
  First obtain an $(CB^{c+1}, B^c)$-network of depth $c+1$ by introducing $CB^{c+1}$ input nodes, 
  denoted by the set $I$, and connect them to $H$ according to a graph with matching up to~$B^c$ in the statement of the proposition. 
  Finally, merge $B$ copies of this graph having the same set $I$ of inputs and disjoint sets of outputs.

  \medskip
  \noindent
  The rearrangeability property follows from the 2 lemmas above.

  Proof that the network has at most $tCD B^{t+1}$ edges. 
  We prove this by induction on $t$. For $t=1$, the network is fully connected and has at most $(CB)\cdot B \le tCDB^{t+1}$ edges. 

  Let $t \ge 2$ and assume that the construction of a $(CB^{t-1}, B^{t-1})$-connector of depth $t-1$ contains at most $(t-1)CDB^t$ edges. 
  The $(CB^t, B^t)$-network consists of $B$ such connectors and $B$ graphs with $CB^t$ left nodes.
  Thus, the total number of edges is at most
  \[
    B \cdot (D \cdot CB^t + (t-1)CDB^t) = tCDB^{t+1}. 
  \]
  
  \noindent
  With exactly the same construction, connectors are obtained.
  The matching game on $(N, N')$-networks is defined in precisely the same way, and using this game, {\em $(N, N')$-connectors} are defined.
  We adapt the 2 lemmas above.
  \begin{itemize}
    \item
      If a graph with sizes $N$ and $N'$ has online matching up to~$K$,
      then its concatenation with an $(N',K)$-connector is an $(N, K)$-connector.
    \item
      The union of $C$ output disjoint $(N, N')$-connectors is an $(N, CN')$-connector.
  \end{itemize}
  For the first item, when Requester selects an input--output pair $(i, o)$,
  this triggers a request for a match $i'$ for $i$ in the graph with online matching, followed by a request to connect the pair $(i',o)$ in the connector. 
  Since there are $K$ outputs, at most $K$ matches are simultaneously needed and therefore both requests can be satisfied.

  The second is easy to understand, since the path finding (or better tree extension) algorithms of separate copies do not interfere.
  Both claims together provide the connectors of the proposition.
\end{proof}

\noindent
To obtain the explicit construction from~ \Cref{cor:explicitConnector}, we use the following explicit lossless expander based on~\cite{guv:j:extractor}.

\begin{theorem}[\cite{lu-oli-zim:c:optimalcode}, Th 18]\label{th:explicitexpander}
  For all $\eps$, $N$, and $K$, there exists an explicit graph with left size $N$, $((1-\eps)D)$-expansion up to~$K$, left degree $D = (\log N)^{O(1)} (\tfrac 1 \eps \log K)^{O(\log \log K)}$, and right size $K \cdot\poly(D\log N)$. 
\end{theorem}

\Cref{cor:explicitConnector} follows by applying this construction to the matching algorithm from
\cref{th:DynamicMatching} applied to the lossless expander obtained from \cref{th:explicitexpander} as follows. Choose $\eps = \tfrac 3 4$ and left size~$N^2$. 
For each $K \le N$, the expander of \cref{th:explicitexpander} satisfies: 
\[
  \max\{\text{left degree},  \frac {\# \text{right set}}{K}\} \;\le\; \poly(\log N \exp O(\log^2 \log K)) \;\le\; \exp(O(\log^2 \log N)),
\]
and this is bounded by $N$ for large $N$. Let $C$ be the right side of the above. 
From this expander, we only use the first $CN^{(c+1)/t}$ of the $N^2$ left nodes and drop the others. 
This yields the graphs satisfying the conditions of \cref{prop:matchig2connector} with $D = \exp(O(\log^2 \log N))$.

\bigskip
\noindent
For non-explicit constructions, we use an expander with smaller degree.
In fact, a random graph has good expansion properties as explained  for example in~\cite[theorem 6.14]{vad:b:pseudorand} or~\cite[appendix C]{bau-zim:j:univcompression}.
The following lossless expander is directly obtained from lemma~\ref{lem:losslessExpander} proven in the appendix. 

\begin{lemma}\label{lem:non-explicitexpander}
  For each $N$ and $K$, there exists a $(\tfrac 3 4 D)$-expander up to~$K$ with left degree $D = \lceil 4\log \tfrac {6N} K\rceil$, left size~$N$ and right size~$16KD$.
\end{lemma}

\noindent
\Cref{cor:nonexplicitConnector} follows from \cref{prop:matchig2connector} and \cref{th:DynamicMatching} instantiated with this expander.

\newcommand{\etalchar}[1]{$^{#1}$}

\appendix

\section{The (slow) online matching algorithm of Feldman, Friedman, and Pippenger}\label{s:slowmatch}

For completeness, we present a special case of~\cite[Proposition 1]{fel-fri-pip:j:networks}. Our proof is based on the original one.
The result implies that if a graph has offline matching up to~$K$, then it has online matching up to~$K$ elements with load $3$.

\begin{theorem}\label{th:ffp}
  If a graph has 1-expansion up to~$K$ and each left set $S$ with $K < \#S \le 2K$ has at least $\#S + K$ neighbors, then the graph has online matching up to~$K$.
\end{theorem}

\begin{corollary}\label{c:ffp}
If a graph $G$ has $1$-expansion up to~$K$, then it has online matching up to~$K$ with load $3$.
\end{corollary}
\begin{proof}
  We modify $G$ by taking 3 clones of each right node. The new graph $G'$ satisfies the hypothesis of~\cref{th:ffp}. Indeed, let $S$ be subset of left nodes with $K < \#S \le 2K$. We partition $S$ into  a set $S_1$ of size $K$ and a set $S_2$ of size $\# S - K \le K$. $S_1$ has at least $2K$ neighbors in the right subset made with the first 2 clones, and $S_2$ has at least $\$S - K$ neighbors in the set made with the third clones. Thus, $S$ has at least $2K + \# S - K = \#S + K$ neighbors.~\cref{th:ffp} implies  that $G'$ has online matching up to~$K$. By merging the 3 clones into the original nodes, it follows that $G$ has online matching with load $3$. 
  \end{proof}

\noindent
We continue with the proof of~\cref{th:ffp}. We start with  2 technical lemmas.

\begin{definition*}\label{def:critical}
  For a set of nodes $S$, let $\N(S)$ be the set of all neighbors of elements in $S$. 
  A left set $S$ is {\em critical} if $\#\N(S) \le \# S$. 
\end{definition*}

\begin{lemma}
  If $A$ and $B$ are critical and $\#\N(A \cap B) \ge \#A \cap B$, then $A \cup B$ is also critical. 
\end{lemma}

\begin{proof}
  We need to bound the quantity $\# \N(A \cup B)$ which equals $\# \N(A) \cup \N(B)$. 
  By the inclusion-exclusion principle this equals
  \[
    = \# \N(A) + \# \N(B) - \# \N(A) \cap \N(B).
  \]
  Since $\N(A \cap B) \subseteq \N(A) \cap \N(B)$ and the assumption of the lemma, this is at most
  \[
    \le \# \N(A) + \#\N(B) - \# A \cap B.  
    \]
   Since $A$ and $B$ are critical, this is at most $\# A + \# B - \#A \cap B = \# A \cup B$.
\end{proof}

\begin{lemma}
 Assume a graph has 1-expansion up to~$K$ and has no critical set $S$ with $K < \# S \le 2K$. 
  Then, for every left node $x$ there exists a right node $y$ such that after deleting $x$ and~$y$, 
  the remaining graph has 1-expansion up to~$K$.
\end{lemma}

\begin{proof}
  A right neighbor $y$ of $x$ is called {\em bad} if after deleting $y$, there exists a left set $S_y$ of size at most $K$ such that $\# \N(S_y) < \# S_y$. 
  Note that $S_y$ is critical, and by the 1-expansion  of the original graph,  $\N(S_y)$ contains~$y$. 
  We show that by iterated application of the above lemma, the set
  \[
    U = \bigcup_{y \text{ is bad}} S_y
  \]
  is critical. Indeed, for each critical set $C$ of size at most $K$, the set $C \cup S_y$ is critical by 1-expansion and the previous lemma. 
  Also this set has cardinality at most $2K$, thus by the assumption this union must have cardinality at most $K$. 

  Note that if all neighbors $y$ of $x$ were bad, then $\N(U \cup \{x\}) = \N(U)$ because $y \in \N(S_y) \subseteq \N(U)$. Thus
  \[
    \# \N(U \cup \{x\}) \le \#U \le \#U \cup \{x\}. 
  \]
  If $\#U < K$, then this violates $1$-expansion, and if $\#U = K$, this violates the assumption about the sizes of critical sets. 
  Hence, at least 1 neighbor of $x$ is not bad and satisfies the conditions of the lemma.
\end{proof}

\begin{proof}[Proof of \cref{th:ffp}.] 
  The online matching strategy maintains a copy of the graph.
  If Requester makes a matching request for a left node $x$, 
  Matcher replies by searching for a right node $y$ that satisfies the condition of the above lemma for the copy graph and adds the edge $(x,y)$ to the matching $M$. In the copy she deletes the nodes $x$ and~$y$. 
  When Requester removes an edge $(x,y)$ from $M$, Matcher restores the nodes $x$ and $y$ in the copy graph.

  It remains to show that in each application of the above lemma, the conditions are satisfied. 
  Note that if Matcher restores the endpoints $x$ and $y$ of an edge, 
  the conditions always remain true, because if $x \not\in S$, then $\#S$ and $\#N(S)$ do not change, and otherwise both values increase by $1$. 

  It remains to show that before any matching request,
  the copy graph has no critical set $S$ with $K < \#S \le 2K$ (and thus the Matcher can apply the lemma and satisfy the request).  Assume to the contrary that there is such an $S$.
  In the original graph, $S$ has at least $\#S + K$ neighbors. 
  When a right neighbor is assigned,  Matcher deletes it from the copy graph. Therefore before any request, the Matcher has deleted from $S$ at most $K-1$ right nodes (since there can be at most $K-1$ active requests), hence,  
  $S$ has at least $\#S + K - (K-1) = \#S + 1$ neighbors, thus it is not critical. 

  Therefore, the conditions of the lemma are always satisfied and the strategy can always proceed by selecting a neighbor~$y$. 
  The theorem is proven.
\end{proof}

\begin{remark*}
  In the matching algorithm from~\cite{fel-fri-pip:j:networks}, the condition on the $1$-expansion up to~$K$ elements is checked using a brute force check over all left sets of size at most~$K$. 
This can be done  in $O({\# L \choose K})$ time. In general, checking whether a graph has $1$-expansion up to~$K$ elements is $\mathsf{coNP}$-complete, see~\cite{bkvpy:j:dispnp}.
However, this hardness result does not exclude algorithms that run in time $\poly(\log \#L)$ for specially chosen graphs.
\end{remark*}

\section{Non-explict lossless expander}

\begin{lemma}\label{lem:losslessExpander}
  For all $\eps \in (0,\tfrac 1 2]$, $N$ and $K \le N$, there exists a graph with 
  left degree $D = \lceil \tfrac 1 \eps \log \tfrac {6N}{K}\rceil$, 
  right size $\tfrac {4} {\eps} DK$, and
  $((1-\eps)D)$-expansion up to~$K$. 
\end{lemma}

\noindent
The proof uses Chernoff's bound in multiplicative form. 
The variant in Hoeffding's original paper~\cite[Theorem 1]{hoe:j:probbound} is used.\footnote{
  In this paper, $\mathbb E X = \mu$ is assumed, but one verifies that the deriviative to $\mu$ of the middle quantity is nonnegative for~$\gamma \ge \mu$.
}

\begin{theorem}[\cite{hoe:j:probbound}]
  Let $X$ be the mean of $n$ independent variables with values in~$[0,1]$. 
  If $\mathbb E X \le \mu \le \gamma < 1$, then
  \[
    \Pr [X \ge \gamma] \;\le\; \left( \frac \mu \gamma \right)^{\gamma n}\left( \frac {1-\mu} {1- \gamma} \right)^{(1-\gamma) n}
\;\le\; \left( \frac \mu \gamma \right)^{\gamma n}\left( \frac 1 {1- \gamma} \right)^{n}. 
  \]
\end{theorem}

\begin{proof}[Proof of lemma~\ref{lem:losslessExpander}.] 
  The probabilistic method is used:
  we prove that if for each left node, $D$ random neighbors are selected, then expansion holds with positive probability. 

  Fix a left set $S$ of size $K'$ with $K' \le K$. 
  Imagine we select random neighbors of nodes in $S$ subsequently. 
  What is the probability that a new neighbor coincides with a previously selected neighbor? 
  Note that if such a collision happens at most $\eps KD$ times, then the set has $(1-\eps)KD$ expansion. 
  For the $i$-th selection, this collision probability is at most $i/\frac{4KD}{\eps}$. 
  Hence, 
  \[
    \mathbb E[ \textnormal{fraction of collisions} ] \;\le\; \frac 1 {K'D} \cdot \frac {\eps}{4KD}  \cdot \sum_{i < K'D} i \;\le\; \frac {\eps} {8} \cdot \frac {K'} K.  
  \]
  By the above Hoeffding bound, the probability that an $\eps$ fraction of collisions happen, is at most 
  \[
    \le \left(\frac {K'}{8K}\right)^{\eps DK'} \left(\frac 1 {1-\eps} \right)^{DK'} \le \left( \frac {K'} {2K} \right)^{\eps DK'}, 
  \]
  where for the right inequality, we used $\tfrac 1 {1-\eps} \le 2^{2\eps}$ if $0 \le \eps \le 1/2$. 

  Fix $K'$. What is the probability that there exists a left set size $K'$ with expansion less than $(1-\eps)D$?
  By the union bound over all $K'$-element sets and the binomial bound ${N \choose K'} \le (eN/K')^{K'}$ with $e=\exp(1)$: 
  \[
    \le \left(\frac {eN} {K'} \right)^{K'} \cdot \left( \frac {K'} {2K} \right)^{\eps DK'} 
    \le \left(\frac {eN} {K} \right)^{K'}\cdot \left( \frac {1} {2} \right)^{\eps DK'}, 
  \]
  where for the right inequality we used $\eps D \ge 1$ and $K'/K \le 1$. 
  Finally, we use $\eps D \ge \log \tfrac {6N} {K}$ to bound this by 
  \[
    \le \left(\frac {eN} {K} \right)^{K'}\cdot \left( \frac {K} {6N} \right)^{K'} \le 2^{-K'}. 
  \]
  Recall that this quantity bounds the probability that some set of fixed size $K'$ exists that does not have enough expansion. 
  It remains to sum this for $K' = 1, 2, \ldots, K$, which is is strictly less than~$1$. 
  Hence, we have proven that with positive probability, a random graph has $((1-\eps)D)$-expansion up to~$K$. 
\end{proof}

\noindent
Note: A similar result with slightly worse constants appears in~\cite[Lemma 4]{bu-mi-ra-ve:c:bitvector}.

\section{Related work on matching}\label{sec:relatedWork}

For more than 4 decades, matching algorithms have been studied, see~\cite{plummer:b:matchingalgorithms1986}, 
and the research still continues, see for example~\cite{behnezhad:c:sublinearmatching}.
We discuss 3 areas in which variants of online matching algorithms are studied. 
The algorithms from theorem~\ref{th:DynamicMatching} and  proposition~\ref{prop:poly_time_matching} combine the constraints of all these areas, 
but these algorithms only work for graphs with large expansion (and for theorem~\ref{th:DynamicMatching}, load is allowed).

\bigskip
\noindent
\textbf{Online matching.}
Let $\mcm(G)$ be the maximum cardinality of a matching in a graph~$G$. 
For some $\alpha$ that is close to $1$, the objective is to maintain a matching of size $\alpha \mcm(G)$ while edges and vertices are added and removed from the graph~$G$. 
Once a match is assigned it may not be revoked. 

A greedy algorithm that maintains a {\em maximal} matching, i.e., a matching that is not a strict subset of another matching, obtains this objective for $\alpha = 1/2$. 
Note that the greedy algorithm in~\cref{cor:TroundMatching}} (presented in the companion paper~\cite{companion-DS}) maintains a maximal matching in the induced subgraph with left nodes that have ``active requests''. 
A similar algorithm is given in the proof of~\cite[corollary 4]{williams:c:dynmatching}.

Perhaps the first paper in this field is~\cite{karp:c:optimalonlinematching}. This paper considers the incremental setting in which left nodes arrive but do not depart. 
They give a probabilistic $(1-1/\exp(1))$-approximation algorithm for bipartite graphs with a perfect matching.\footnote{The adversary is oblivious, i.e., the moves of the adversary are fixed before the randomness of the algorithm is fixed.} More recently, this was improved to an $(1-1/D)$-approximation for regular graphs with degree~$D$. Unfortunately, the runtime of this algorithm is polynomial in~$N$,~\cite{cohen:c:onlinematching}.


\bigskip
\noindent
\textbf{Dynamic matching.}
The objective is again to maintain a matching of at least $\alpha\mcm$, but now matches are allowed to be {\em revoked}. 
The aim is to minimize the runtime and it is also important to have few revocations. See~\cite{hen-kha-pau-sch:c:dynamicmatching,BehnezhadGhafari2024FOCS} for references to recent results.

In~\cite[theorem 2]{williams:c:dynmatching} an algorithm is given for general graphs that are not necessarily bipartite.  
A $(1/2)$-approximation is given in which the worst case number of revocations for each assigned match is 1, 
and the amortized runtime is $O(D)$ in the word-ram model, where $D$ is the average degree of the graph.
Note that this algorithm is almost irrevocable. 
If it could be made irrevocable, we would obtain a stronger version of theorem~\ref{th:DynamicMatching} that does not require lossless expansion:
$(1/c)$-expansion up to $K$ implies $\poly(\log N)$-time matching with load $O(c\log N)$ up to~$K$. 
Unfortunately, such an improvement is unlikely, because it would contradict 2 popular conjectures:
the ``online matrix-vector multiplication conjecture'' and ``triangle detection conjecture'', see~\cite[theorem 1]{williams:c:dynmatching}. 



\bigskip
\noindent
\textbf{Load balancing with restricted assignment.}
In this task, there are $M$ servers and tasks arrive with a duration and a subset of servers that can perform the task. 
When the tasks arrive, a server needs to be selected immediately. For the full duration of the task, the servers' load is increased by 1. 
Usually, it is not allowed to reassign a task to a different server. The aim is to minimize the maximal load of a server. 
Two types of assignment algorithms are studied, depending on whether the input contains the duration of the task. 
Our game corresponds to the variant in which the duration is not given.

Given a sequence of clients, the performance of an assignment algorithm is the maximum of the load over all machines and over time.
The aim is to minimize the competitive ratio: the ratio of this performance to the performance of an optimal offline algorithm that is given the full sequence of tasks and there durations at once. 
We refer to~\cite{azar:c:loadbalancingsurvey} for more background.

For this model, for every deterministic assignment algorithm, there exists a sequence of tasks of length $\poly(M)$ in which the competitive ratio is at least $\lfloor \sqrt{2M} \rfloor$, 
where $M$ is the number of servers, see~\cite{azar:j:onlineloadbalancing, ma:j:loadbalancing}. 
There exists an algorithm that guarantees a competitive ratio of at most $2\sqrt{M}+1$,~\cite{azar:c:loadbalancingRobinHood}. 
The algorithm from the proof of theorem~\ref{th:DynamicMatching} obtains a competitive ratio $O(\log N)$, 
which is typically exponentially smaller than $2\sqrt{M}+1$, 
but this algorithm only works for graphs that are lossless expanders.

\end{document}